\documentclass[letter]{aa} 

\usepackage{graphicx}
\usepackage{txfonts}
\usepackage{natbib,twoopt}
\usepackage[breaklinks=true]{hyperref} 
\bibpunct{(}{)}{;}{a}{}{,} 
\makeatletter
\newcommandtwoopt{\citeads}[3][][]{\href{http://adsabs.harvard.edu/abs/#3}%
{\def\hyper@linkstart##1##2{}%
\let\hyper@linkend\@empty\citealp[#1][#2]{#3}}}
\newcommandtwoopt{\citepads}[3][][]{\href{http://adsabs.harvard.edu/abs/#3}%
{\def\hyper@linkstart##1##2{}%
\let\hyper@linkend\@empty\citep[#1][#2]{#3}}}
\newcommandtwoopt{\citetads}[3][][]{\href{http://adsabs.harvard.edu/abs/#3}%
{\def\hyper@linkstart##1##2{}%
\let\hyper@linkend\@empty\citet[#1][#2]{#3}}}
\newcommandtwoopt{\citeyearads}[3][][]%
{\href{http://adsabs.harvard.edu/abs/#3}
{\def\hyper@linkstart##1##2{}%
\let\hyper@linkend\@empty\citeyear[#1][#2]{#3}}}
\makeatother
\begin{document} 

   \title{Non-azimuthal linear polarization in protoplanetary disks
}

\author{H. Canovas\inst{1,5}
   F. M\'enard\inst{2},
   J. de Boer\inst{3},
   C. Pinte\inst{2},
   H.  Avenhaus\inst{4,5},   
   M.~R. Schreiber\inst{1,5}
}

\institute{ 
Departamento de F\'isica y Astronom\'ia, Universidad de Valpara\'iso,
Valpara\'iso, Chile, \email{hector.canovas@dfa.uv.cl} 
\and
   UMI-FCA, CNRS/INSU, France (UMI 3386), and Dept. de Astronom\'{\i}a, Universidad de Chile, Santiago, Chile. 
\and
 Leiden Observatory, Universiteit Leiden, P.O. Box 9513, 2300 RA Leiden, The Netherlands.
\and
Departamento de Astronom\'ia, Universidad de Chile, Casilla 36-D, Santiago, Chile.
\and
  Millennium Nucleus ``Protoplanetary Disks in ALMA Early Science"
}

\date{\today}

 
  \abstract
   {Several studies discussing imaging polarimetry observations of protoplanetary disks use the so-called radial Stokes parameters
   $Q_\phi$ and $U_\phi$ to discuss the results. This approach has the advantage of providing a direct measure of the noise in the
   polarized images under the assumption that the polarization is azimuthal only, i.e., perpendicular to the direction towards the illuminating
   source. However, a detailed study of the validity of this assumption is currently missing.
   We aim to test whether departures from azimuthal polarization can naturally be produced by scattering processes in optically
    thick protoplanetary disks at near infrared wavelengths.
   We use the radiative transfer code MCFOST to create a generic model of a transition disk using  different grain size distributions
   and dust masses.  From these models we generate synthetic polarized images at $2.2\,\mu$m.
   We find that even for moderate inclinations (e.g., $i = 40\degr$), multiple scattering
   alone can produce significant (up to $\sim4.5\%$ of the $Q_\phi$ image, peak-to-peak) non-azimuthal polarization reflected in the
   $U_\phi$ images. We also find that different grain populations can naturally produce
   radial polarization (i.e., negative values in the $Q_\phi$ images).
   Despite the simplifications of the models, our results suggest that caution is recommended
   when interpreting polarized images by only analyzing the $Q_\phi$ and $U_\phi$ images. We find that there can be astrophysical
   signal in the $U_\phi$ images and negative values in the $Q_\phi$ images, which indicate departures from azimuthal polarization.
   If significant signal is detected in the $U_\phi$ images, we recommend to check the standard $Q$ and $U$ images to look for 
   departures from azimuthal polarization. On the positive side, signal in the $U_\phi$ images once all instrumental and data-reduction
   artifacts have been corrected for means that there is more information to be extracted regarding the dust population and particle density.}

\titlerunning{Non-azimuthal linear polarization in protoplanetary disks}
\authorrunning{Canovas et al.}

   \maketitle
%

\section{Introduction}
With the advent of high contrast imaging polarimeters such as NaCo/VLT \citep{2003SPIE.4841..944L},
HiCiao/Subaru \citep{2006SPIE.6269E..0VT}, GPI/Gemini South \citep{2008SPIE.7015E..18M}, and
SPHERE/VLT \citep{2008SPIE.7014E..18B}, polarized differential imaging (PDI) has become a standard
tool to directly image and analyze the dusty circumstellar environments around young and evolved stars
\citep[e.g.][]{2012A&amp;A...543A..70C, 2015A&amp;A...578L...1C, 2012A&A...539A..56J, 2013A&amp;A...560A.105G,
2013A&A...554A..15M, 2014A&amp;A...572A...3J}. In disks, PDI have proven powerful to show a variety of features like
spiral arms, bright inner rings and/or disk asymmetries revealing complex morphologies. However, the yield
products of this technique are generally polarized intensity ($P_I$) maps, and rarely intensity and fractional
polarization ($P$) maps. The optical properties of dust grains and their dependence  on composition and scattering
angle have a significant impact on the polarization of the scattered light \citep[e.g., ][]{2010A&A...518A..63M, 2012A&amp;A...537A..75M}.
Multiple scattering and optical thickness can also modify the observed polarization. Therefore, $P_I$
maps may not always trace accurately the underlying surface brightness distribution, and even less so
the surface density distribution \citep[as in AB Aurigae, see][]{2009ApJ...707L.132P}.

Recently it has become popular to analyze linearly polarized images of protoplanetary disks using the so-called 
radial-Stokes formalism \citep{2006A&amp;A...452..657S, 2014ApJ...781...87A}. This is extremely useful because
it allows to quantify and control the calibration errors potentially to high degrees of accuracy. This formalism relies
on the assumption that the linear polarization in disks is perpendicular to the scattering plane defined by the light source
(star), the scattering particle, and the observer, i.e., it is purely azimuthal. Any non-azimuthal polarization is associated
to noise and/or calibration errors.
While this is valid when \textit{single} scattering dominates in nearly face-on disks, the assumption may
break down for higher disk inclinations and for peculiar dust properties \citep[as for those of comets, e.g., ][]{2004AJ....127.2398K},
when grains are aligned \citep{1996ASPC...97...63M, 2002A&A...385..365W}, and/or when disks are optically thick
and multiple scattering occurs \citep{1988ApJ...326..334B}. In this letter we explore some of these deviations and quantify
their impact on the use of the radial-Stokes formalism at near-infrared (NIR) wavelengths.
\section{Modeling}

The goal of this study is to demonstrate that departures from centro-symmetric azimuthal polarization may occur
in protoplanetary disks for several cases. To prove this we build a simple but realistic model of a transition disk
using the 3D radiative transfer code MCFOST \citep{2006A&amp;A...459..797P, 2009A&A...498..967P}. The radial
structure of the model is similar to that found in the well studied disks RX\,J1633.9-2442 and Sz\,91
\citep{2012ApJ...752...75C, 2015ApJ...805...21C}: the innermost region is completely devoid of dust, followed by a
dust-depleted region extending up to the cavity radius (``Region 1''), and the outer disk (``Region 2'').

\subsection{Disk structure}

The exact choice of the disk geometry and stellar type has a minimal impact on our study of the polarization when
compared to the effect of dust grain properties and/or optical thickness. We assume the disk to be at the distance
of the Taurus star forming region (140 pc), with central star with mass of $1M_{\sun}$, effective temperature of $T_{\star} = 4300$\,K, 
and radius of $R_\star = 2.6\,R_{\sun}$. This roughly corresponds to a K6 star of age $\sim1$\,Myr \citep{2000A&amp;A...358..593S}.
The disk has an inner cavity of 40\,au in radius ($R_\mathrm{cav}$) containing $10^{-9}$ $M_\sun$ of dust (optically thin at NIR),
with the innermost 10 au totally devoid of dust. The surface density distribution of this inner disk is governed by a
power-law $\Sigma (r) = \Sigma_{100} \left (r / 100\,\mathrm{au} \right)^{-p}$ where $\Sigma_{100}$ is the surface density
at $r = 100$ au. The surface density distribution of the outer disk is described by a tapered-edge profile
\begin{equation}
\Sigma (r) = \Sigma_C \, r^{-\gamma} \, \mathrm{exp} \left [ - \left (\frac{r}{R_{\rm{C}}} \right) ^{2-\gamma} \right ],
\label{eq:eq1}
\end{equation}
where $\Sigma_C$ is the surface density at the characteristic radius $R_{C}$, and $\gamma$ corresponds
to the viscosity power-law index in accretion disk theory \citep[$\nu\propto R^{\gamma}$,][]{1998ApJ...495..385H}.
The scale height is described assuming that the dust follows a Gaussian vertical density profile in both
the inner and outer disks $H(r) = H_{100} (r/100\,\mathrm{au})^\psi$, where $H_{100}$ is the scale height at
$r=100$\,au and $\psi$ is the flaring parameter of the disk. The optical depth is changed by using
different outer disk's dust masses.
\subsection{Dust properties}
Models show that the polarizability curve of different grain types is positive (polarization perpendicular to
the scattering plane) in most cases, but it  can have a negative branch producing polarization parallel to the scattering plane
\citep[i..e, radial polarization, see ][]{2009ApJ...707L.132P, 2013A&amp;A...556A.123C, 2014A&A...568A.103K}.
Although the detailed composition of disk grains is difficult to constrain, observational evidence
indicates that cosmic dust particles are mostly composed of silicates and, in lower quantity, carbonaceous
particles \citep{1984ApJ...285...89D, 2003ARA&amp;A..41..241D}. 
For our modeling we use
grain particles composed of a mixture in volume of $70\%$ astronomical silicates \citep{1984ApJ...285...89D}
and $30\%$ amorphous carbon particles \citep{1997A&amp;A...323..566L} following a standard power-law
size distribution $\mathrm{d}n(a) \propto a^{-3.5} \mathrm{d}a$ using different values for the minimum
and maximum grain sizes. The full scattering matrix of the dust populations is computed with Mie theory
\citep{mie_1908}, using two different methods: assuming distributions of 1) compact homogeneous spheres, and
2) hollow spheres \citep[DHS,][]{2005A&A...432..909M} with maximum volume fraction $f_{\max} = 0.8$.
The minimum and maximum grain sizes in Region 1 is fixed and these grains are compact. For the outer disk,
we explore different values of porosity and minimum and maximum grain sizes. The optical indices are derived
using effective medium theory (using Bruggeman's rule). The model's parameters are summarized in Table~\ref{tab:tab1}. 
\begin{table}[t]
\center
\caption{Model parameters. Subscripts refer to the two regions of the disk (``1'' refers to the inner disk,
``2'' refers to the outer disk).}
\begin{tabular}{ c  c  c}
\hline\hline
Parameter                                         &       Symbol                                    &         Value(s)\\
\hline
\hline
Inner radius                                      &         $R_\mathrm{in}$                   & $10$ au                                                 \\
Cavity radius                                    &         $R_\mathrm{cav}$                 & $40$ au                                                 \\
Dust mass                                        &         $M_\mathrm{dust_1}$           &$1\times10^{-9}$ $M_\sun$                   \\
Surface density index                       &         $p$                                        & $-1$                                                        \\
Scale height                                     &         $H_{100_1}$                          &$10$ au                                                   \\
Flaring index                                    &        $\psi_{1}$                                & $1.15$                                                     \\	
Min. grain size                                  &        $a_{\rm{min_1}}$                    & $3\,\mu$m                                             \\
Max. grain size                                 &        $a_{\rm{max_1}}$                   &$20\,\mu$m                                             \\
Porosity                                           &        $\mathcal{P}$                          & 0 [$\%$]                                                  \\
\hline
Characteristic radius                        &         $R_{C}$                                 & $100$ au                                                 \\
Dust mass                                        &         $M_\mathrm{dust_2}$           &$0.5\times [10^{-8}, 10^{-5}, 10^{-4}]\,M_\sun$             \\
Surface density index                      &          $\gamma$                             & $0.3$                                                       \\
Scale height                                     &         $H_{100_2}$                          &$10$ au                                                    \\
Flaring index                                    &        $\psi_{2}$                                & $1.15$                                                     \\
Min. grain size                                 &        $a_{\rm{min_2}}$                    & $0.03,0.2,5,10\,\mu$m                            \\
Max. grain size                                &        $a_{\rm{max_2}}$                   & $1000, 1500, 2000\,\mu$m                       \\
Porosity                                           &        $\mathcal{P}$                          & 0, 20, 30, 40 [$\%$]                                \\
\hline
\end{tabular}
\label{tab:tab1}
\end{table}

\subsection{Synthetic scattered light images}
We use MCFOST to generate synthetic images (Stokes I, Q and U maps) at 2.2$\,\mu$m
of the disk models for several inclinations ranging from pole-on to edge-on. To facilitate
the comparison with real observations these images are projected into a grid with pixel size of
$0\farcs027\,\mathrm{px}^{-1}$ (equal to the scale of the \textit{S27} camera of NaCo/VLT)
and convolved with a 2.5-px width ($0\farcs067$) gaussian point spread function (PSF). After convolution, the polarized
intensity image is constructed as $P_{I} = \sqrt{Q^2 + U^2}$. The radial-Stokes parameters\footnote{
The nomenclature/symbols for these quantities is not standardized. For consistency with previous studies
we use the term ``radial-Stokes''.}
are defined as
\begin{eqnarray}
Q_\phi & =&  +Q \cos (2\phi) + U \sin (2\phi) \\
U_\phi & = & -Q \sin (2\phi) + U \cos (2\phi)
\label{eq:eq4}
\end{eqnarray}
where
\begin{equation}
\phi = \arctan \frac{x - x_0}{y - y_0} + \theta,
\label{eq:eq5}
\end{equation}
$x$ and $y$ are the coordinates in the image with respect to the star ($x_0$, $y_0$) and $\theta$ is the offset
created by instrumental polarization (zero in our models).

\begin{figure}[t!]
  \centering \includegraphics[width=\columnwidth, trim = 10 10 50 10]{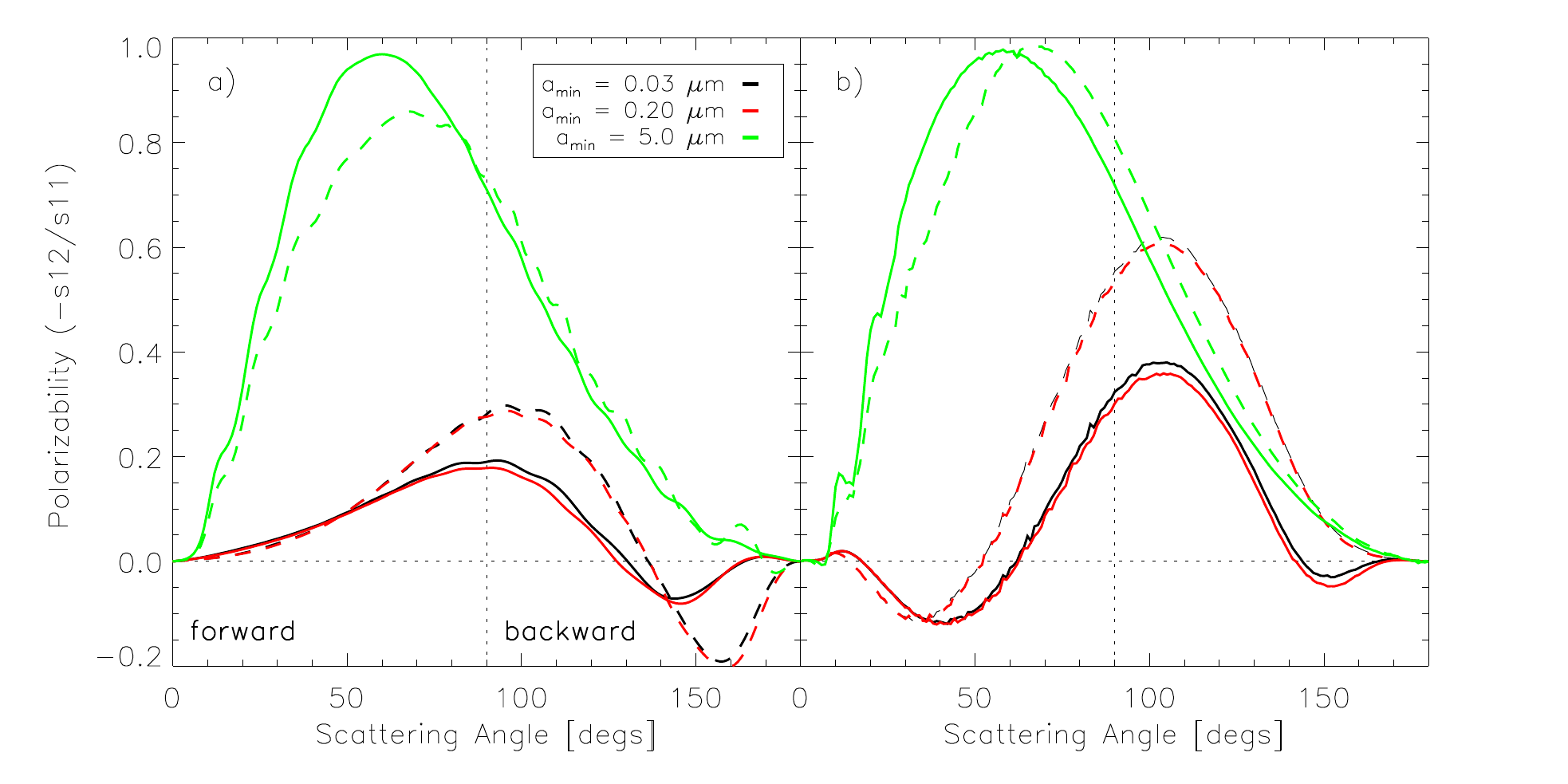}
  \caption{ a) Polarizability curves computed at $2.2\,\mu$m using solid spheres for $a_\mathrm{max} = 1$ mm
  and different values of $a_\mathrm{min}$ (see legend). Solid (dashed) lines represent $0\%$ ($30\%$) porous
  grains. Negative values indicate radial polarization. b) Same as a) using DHS theory with $f_\mathrm{max} = 0.8$.}
   \label{fig:fig1}
\end{figure}

\section{Results and discussion}

\subsection{The effect of dust properties}

The angle at which incident light is scattered and the polarized flux that is produced strongly depends
on the detailed properties of the scattering dust particles. The polarizability curves of different grain distributions
prove this dependency for \textit{single} scattering of incident \textit{unpolarized} light. Fig.~\ref{fig:fig1} shows
that different grain populations can rotate the polarization plane from perpendicular (positive branch) to parallel
(negative branch) with respect to the scattering plane. The first case produces azimuthal polarization patterns
(positive signal in the $Q_{\phi}$ images) while the second case produces radial polarization (negative values
in $Q_{\phi}$). In our models, both the DHS and the solid spheres produce significant amounts of radial polarization
for $a_\mathrm{min} < 5\,\mu$m. Using solid spheres, radial polarization is expected only for angles $\gtrsim130\degr$
(backward scattering). For a protoplanetary disk this would correspond to scattering on its far side at 
inclinations $i \gtrsim (40\degr + \alpha)$, where $\alpha$ is the flaring angle of the disk. On the other hand
DHS predicts radial polarization for scattering angles $\lesssim70\degr$ (forward scattering) and $140\degr-160\degr$
(backward scattering), noting that different values of $f_\mathrm{max}$ can remove the
negative branch for forward scattering angles \citep{2005A&A...432..909M}. 

Radial polarization at large scattering angles (backward scattering) has been observed at optical and NIR wavelengths
in a number of comets \citep{2004AJ....127.2398K, 2011AJ....141..181W}, and dust grains in protoplanetary disks can
have similar compositions to those found in comets \citep{2003A&A...401..577B}. Experimental measurements in the
optical show that ensembles of arbitrarily shaped dust aggregates also produce radial polarization
patterns at large scattering angles \citep{2000A&A...360..777M, 2006A&amp;A...446..525M, 2007A&A...470..377V}.
At optical wavelengths models show that porous grains can also produce radial polarization in selected regions
of protoplanetary disks at large inclinations \citep{2014A&A...568A.103K}. Therefore, dust properties alone can produce
radial polarization in particular for backward scattering. 

\subsection{The effect of multiple scattering}

Multiple scattering of previously polarized light may lead to complex polarization patterns. In particular, the polarization
plane of the last scattering event could have an orientation that is neither parallel nor perpendicular to the scattering
plane of the first scattering event. This would be reflected in the $U_{\phi}$ images. To describe this case
while minimizing the impact of the grain properties, we focus in models having $a_{min} \geq 5\mu$m, i.e., models
with polarizability curves without negative branches (as the green curve in Fig.~\ref{fig:fig1}). The contribution of multiple
scattering can be estimated by comparing models with the same grain populations but different masses (i.e.,
different number of scattering particles), as higher masses will produce more scattering events.

Three representative cases are shown in Fig.~\ref{fig:fig2}. For each model, the images are normalized with respect
to the $P_I$ image to facilitate a comparison between the $P_I, Q_\phi$, and $U_\phi$ images. All models are computed
using $\mathcal{P} = 30\%$, $a_{min, max} = 5,1000\,\mu$m, $i = 40\degr$, and DHS. From top to bottom, the outer disk's
dust masses are $M_{\mathrm{dust_2}} = 0.5\times [10^{-8}, 10^{-5}, 10^{-4}]\,M_\sun$, corresponding to integrated optical
depths at 2.2 $\mu$m along the disk's mid plane of $\tau_{2.2} = [10^{-3}, 8.3, 83.0]$. By construction there is
no instrumental polarization in our models and the numerical noise from the Monte Carlo simulation is at most
$0.4\%$ of the $P_I$ image (peak to peak, see the optically thin model in top row of Fig.\ref{fig:fig2}).

We find that multiple scattering can naturally produce observable signatures in the unconvolved $U_{\phi}$ images even in
disks with moderate inclination ($i = 40\degr$). This is reflected in the $U_\phi$ images in Fig.~\ref{fig:fig2}, which can have maximum
values up to $\sim4.5\%$ of the $P_I$ image (peak-to-peak). In all our models the signal in the $U_\phi$ image relative to the $P_I$
image increases with inclination
and with mass. The $Q_\phi$ images, on the other hand, are mostly insensitive to this effect. Both the DHS and compact spheres
methods produce the same trends. Additionally we find that increasing the porosity slightly increases the signal in $U_\phi$, but this
feature becomes undetectable after convolution with the PSF. Instrumental effects like PSF convolution
can remove these trends and/or even create artificial signatures in the $U_\phi$ images.  

Models based on Mie theory are an attempt to represent the complex, fluffy shapes observed in the dust
particles of the interstellar medium. For example, detailed, multi wavelength studies of the optically
thin debris disks HD\,181327 and HR\,4796 A \citep{2012A&A...539A..17L, 2015A&amp;A...577A..57M}
show that  neither by using solid spheres nor DHS is it possible to reproduce the observed scattering properties
of their dust grains. While it is clear that a more elaborate theory is needed to successfully
reproduce the scattering properties of the grains, it is noteworthy that both the compact
spheres and the DHS methods predict radial polarization features (i.e., negatives in the $Q_\phi$ image)
for several types of grain distributions, and an increment in signal in the $U_\phi$ images with
multiple scattering.

\section{Conclusions}

We investigated the conditions for which non azimuthal polarizations can be produced at significant levels in dusty
circumstellar environments, and in particular in young protoplanetary disks. By doing so we investigated as well the
validity of the assumption currently made when reducing PDI data using the radial Stokes parameters, namely that
the linear polarization in disks is strictly azimuthal and any non-azimuthal polarization is produced by noise and/or
instrumental effects.

\begin{figure*}[t!]
  \centering \includegraphics[width=0.8\textwidth, trim = 0 90 0 0]{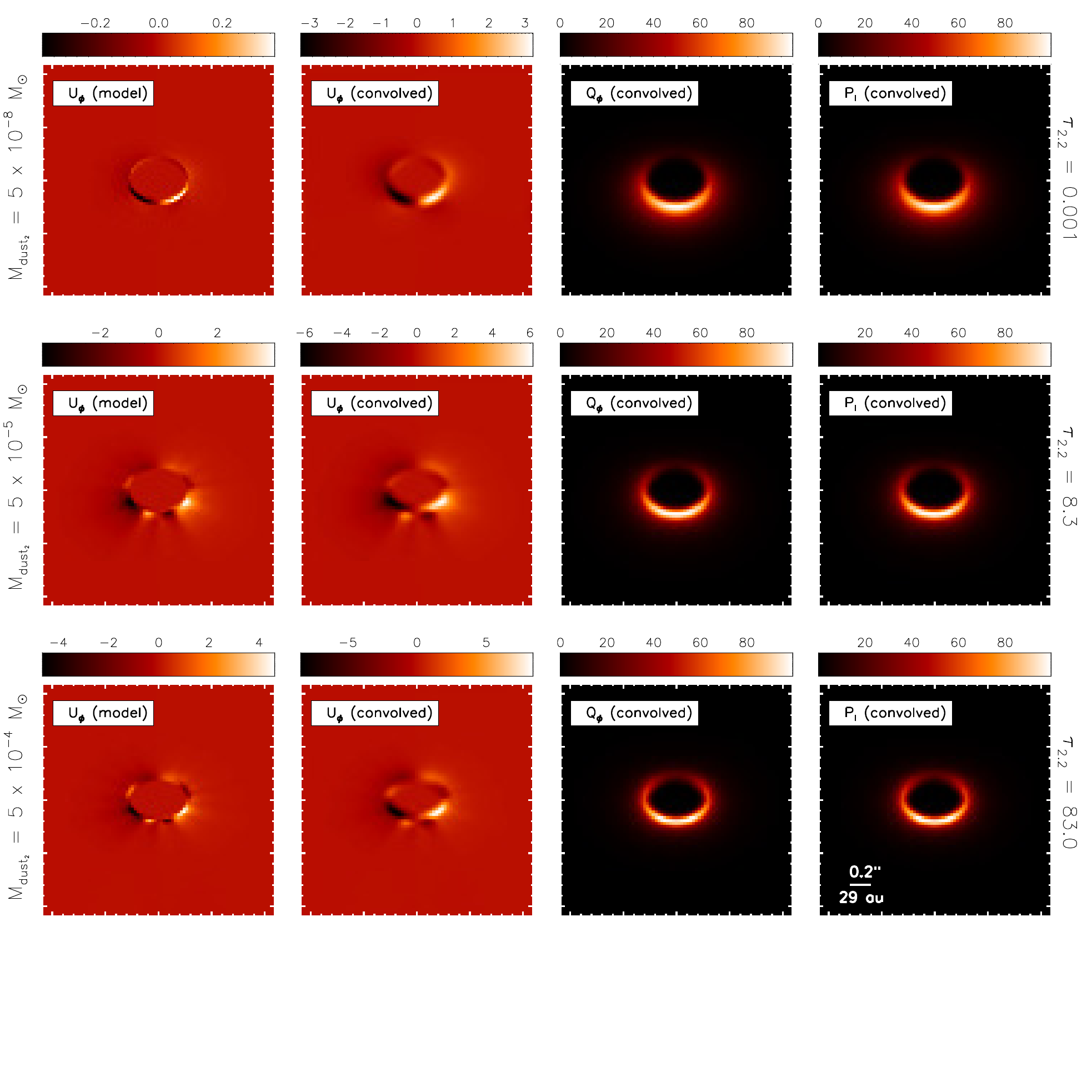}
  \caption{From top to bottom: synthetic images at $2.2\,\mu$m for three models ($i = 40\degr$) with outer disk masses of
  $M_{\mathrm{dust_2}} = 5\times[10^{-8}, 10^{-5}, 10^{-4}]\,M_\sun$. From left to right, unconvolved $U_\phi$ images, and
  convolved $U_\phi, Q_\phi, P_I$. All models were computed using the DHS method and share the same grain populations
  and porosity (see text). For each row, the images are scaled to the $P_I$ image with the maximum value of $P_I$
  being 100. Neither noise nor instrumental polarization are included in these simulations, highlighting the relative
  changes in the $U_\phi$ and $Q_\phi$ images with multiple scattering.}
   \label{fig:fig2}
\end{figure*}

Protoplanetary disks are usually optically thick at NIR wavelengths and multiple light scattering alone, or in other words the
scattering of polarized light, can readily produce detectable signatures in the $U_{\phi}$ images. These signatures increase
with inclination angle and optical thickness, and this trend holds for compact and porous grains. In the disk models discussed
above the signal in $U_\phi$ can reach up to $\sim4.5\%$ of the $P_I$ (unconvolved, peak-to-peak) for a disk inclined by $40\degr$.
This value goes up to $50\%$ for an inclination of 70\degr for the most massive disk shown in Fig.~\ref{fig:fig2}.
Furthermore, dust properties alone may also lead to violation of the strictly azimuthal linear polarization assumption. Astronomical
observations, laboratory measurements, and models indicate that it is quite common in nature to find dust chemical compositions,
shapes and size distributions that will produce azimuthally centrosymmetric polarization for most scattering angles but that
will rotate the polarization plane by $90\degr$ for large scattering angles. This will produce radial polarization resulting in
negative signal in the $Q_\phi$ images. This is often the case in comets and similarities are found, at least in terms of chemical
composition (spectroscopic signatures), between cometary dust and  protoplanetary disk dust. 

We have demonstrated that genuine astrophysical signal in the $U_\phi$ images is expected for a number of very plausible
configurations in protoplanetary disks and other circumstellar environments. A detailed analysis of these scenarios and their
signatures in the $Q_\phi$ and $U_\phi$ images is beyond the scope of this letter but we stress that caution is in order when
assuming that linear polarization is purely azimuthally centrosymmetric in protoplanetary disks, in particular for disks with
inclinations $i\gtrsim 40\degr$. At lower inclinations our models show that the $U_{\phi}$ images usually contain
low signals (at least for the disk models tested here), and it is probably safe to use the radial formalism. 
If significant signal is found in the $U_{\phi}$ images, the formalism may remain relevant for disk detection experiments, 
but for quantitative studies (surface brightness, colors, polarization levels) we recommend to analyze as well the ``traditional''
$Q$ and $U$ images. On the positive side, signal in the $U_\phi$ images once all instrumental and data-reduction artifacts have been
corrected for implies that there is more information to be extracted regarding the dust population and particle density.

\begin{acknowledgements}
        We thank the referee for his/her useful comments.
        We thank M. Min for carefully reading the manuscript.
        This research was funded by the Millennium Science Initiative, Chilean Ministry of Economy, Nucleus
        RC130007. HC acknowledges support from ALMA/CONICYT (grants 31100025 and 31130027).
        MRS acknowledges support from FONDECYT grant 1141269. H.A. acknowledges support from
        FONDECYT 2015 Postdoctoral Grant 3150643. FMe and CP acknowledge funding from the
        EU FP7-2011 under Grant Agreement No 284405. 
\end{acknowledgements}

%
%


\bibliographystyle{aa.bst}      
\bibliography{pol.bib}         

\end{document}